ANALYSIS OF SUPER-KAMIOKANDE 5-DAY MEASUREMENTS OF THE SOLAR NEUTRINO FLUX

P.A. Sturrock
Center for Space Science and Astrophysics, Varian 302,
Stanford University, Stanford, CA 94305

ABSTRACT

Data in 5-day bins, recently released by the Super-Kamiodande Consortium, has been analyzed by a likelihood procedure that has certain advantages over the Lomb-Scargle procedure used by the consortium. The two most prominent peaks in the power spectrum of the 10-day data were at 9.42 $y^{-1}$ and 26.57 $y^{-1}$, and it was clear that one was an alias of the other caused by the regularity of the binning. There were reasons to believe that the 9.42 $y^{-1}$ peak was an alias of the 26.57 $y^{-1}$ peak, but analysis of the 5-day data makes it clear that the reverse is the case. In addition to a strong peak near 9.42 $y^{-1}$, we find peaks at 43.72 $y^{-1}$ and at 39.28 $y^{-1}$. After comparing this analysis with a power-spectrum analysis of magnetic-field data, we suggest that these three peaks may be attributed to a harmonic of the solar rotation rate and to an r-mode oscillation with spherical harmonic indices l = 2, m = 2.

*Subject headings:* neutrinos – Sun: interior – Sun: particle emission – Sun: rotation

1. INTRODUCTION

In August 2002, the Super-Kamiokande Consortium (Fukuda et al. 2001, 2002) made available a dataset suitable for time-series analysis, comprising 184 flux and error estimates drawn from the time interval June 1996 to July 2001 (Smy 2002). Each pair of measurements was extracted from measurements acquired over an interval of about 10 days. The "binning" was highly regular: a power-spectrum analysis of the timing has a huge peak at 35.98 $y^{-1}$ (period 10.15 days), where frequencies are measured in cycles per year. As a result, it is inevitable that any power-spectrum analysis of this dataset will be severely contaminated by aliasing. Our analysis by a likelihood method (Sturrock, Walther, & Wheatland 1997) focused on a prominent pair of peaks, one at 9.42 $y^{-1}$ and the other at 26.57 $y^{-1}$ (Sturrock & Caldwell 2003; Sturrock 2003a). Since these frequencies sum to 35.99 $y^{-1}$, it was obvious that one was an alias of the other; the available evidence seemed to support the interpretation that the primary oscillation was that at 26.57 $y^{-1}$, and that the peak at 9.42 was an alias. Milstajn (2003) analyzed the same dataset by means of the Lomb-Scargle procedure (Lomb 1976; Scargle 1982, 1989), referring each measurement to the mid-time of each bin. This analysis also yielded evidence for an oscillation at 26.57 $y^{-1}$. In all three articles, it was proposed that this oscillation could be attributed to rotational modulation of an m = 2 structure within the Sun, since 26.57 $y^{-1}$ falls in the band 26.36 – 27.66 $y^{-1}$, formed from twice the range of synodic rotation frequencies of an equatorial section of the Sun for normalized radius larger than 0.1 (Schou et al. 1998).

Nakahata (2003), of the Super-Kamiokande Consortium, also carried out a Lomb-Scargle analysis of the above 10-day data set, assigning each measurement to the mid-time of each bin, and found the peak in the spectrum at 26.57 $y^{-1}$. However, Nakahata also analyzed the dataset in terms of the "mean live time" [representing the "center of gravity" of the non-uniform data-acquisition "window" (see equation (4))], for which a Lomb-Scargle analysis yields a reduced power at 26.57 $y^{-1}$, leading Nakahata to suggest that the peak is spurious.

More recently, the Super-Kamiokande Consortium has subdivided the original bins to produce a dataset with 368 bins that have a mean duration of about 5 days (Yoo et al. 2003). Once again, the bin timing is highly regular: the power spectrum of the timing now has a huge peak at 72.01 $y^{-1}$, corresponding to a period of 5.07 days. The Super-Kamiokande publication contains the results of a Lomb-Scargle analysis of this new dataset using the mean live time: there is no peak at or near 26.57 $y^{-1}$ and the authors find no other significant feature in the power spectrum, leading them to conclude that there is no evidence for periodicity of the solar neutrino flux as measured by the Super-Kamiokande experiment.

Since the neutrino dataset is so sparse, it is essential to use as much information as possible. The first point to note, in comparing the usual Lomb-Scargle and the likelihood methods, is that the former does not take account of error estimates (although Scargle (1989) shows how this can be done), but the latter easily can do so.

The second point to note is that the likelihood approach can incorporate information concerning the data acquisition schedule. The available dataset lists the start time $t_{s,n}$ and end time $t_{e,n}$ of each bin, enumerated by n = 1,…, N. The acquisition is not uniform, so the flux estimate $g_n$ depends on the neutrino flux f(t) and on a "weighting function" or "time window function" $W_n(t)$ by an equation of the form

$$g_n = K \int_{t_{s,n}}^{t_{e,n}} dt\, W_n(t) f(t) . \qquad (1)$$

The error estimates will similarly depend upon the time window function. In the likelihood approach (Sturrock, Walther & Wheatland 1997), it is necessary to estimate the measurements $G_n$ that are to be expected from a model of the flux $F(t,\alpha,\beta,...)$, where $\alpha,\beta,...$ are parameters to be estimated. These estimates are in principle given by

$$G_n(\alpha,\beta,...) = K \int_{t_{s,n}}^{t_{e,n}} dt\, W_n(t) F(t,\alpha,\beta,...) . \qquad (2)$$

However, it may not be practical to determine each time window function in detail, in which case we need to adopt a model for these functions based on the available data. When we are given only the start time and end time of each bin, a reasonable model is the "boxcar" model for which W is constant over the bin. This is the assumption that we made in earlier articles in which we analyzed the first dataset that listed only the start time and end time (Sturrock & Caldwell 2003, Sturrock 2003a). Since later datasets list also the mean live time of each bin, we need to adopt a model that takes account of all three times (start time, end time, and mean live time), and we propose such a function in Section 2.

It may help to clarify the relationship between the Lomb-Scargle method and the likelihood method if we note that the former is equivalent to a likelihood analysis that (a) ignores experimental error estimates and (b) adopts a delta-function model for the window function. The two Lomb-Scargle analyses that have been made by the Super-Kamiokande consortium correspond to

$$W_n(t) = \delta(t - t_{mid,n}) \quad and \quad W_n(t) = \delta(t - t_{mlt,n}), \tag{3}$$

in which $t_{mid,n}$ and $t_{mlt,n}$ denote the mid-time and the mean live time, respectively, of each bin. Figure 1 shows the power spectrum for the frequency range 0 – 50, computed by the likelihood method, adopting a narrow time window function (to simulate a delta function) centered on the mean live time and a constant error term. As we expect, this power spectrum is very close to the power spectrum computed by Yoo et al. (2003). As noted by Yoo et al., the peak at 26.57, which was prominent in analyses of the 10-day data, has virtually disappeared.

A review of earlier evidence for variability of the solar neutrino flux and a discussion of the implications of these studies for particle physics are discussed in Caldwell & Sturrock (2003).

## 2. EXTENDED TIME WINDOW FUNCTIONS

Clearly, a delta-function is not a desirable representation of a time window function the length of which is comparable with the period of some of the oscillations being investigated. We therefore repeat our recent analysis (Sturrock 2003a) for the 5-day data set. This analysis uses the start time and end time of each bin, effectively adopting a uniform ("boxcar") time window function for each bin. The resulting power spectrum for the frequency range 0 to 50 is given in Figure 2. Where there was a peak at 9.42 $y^{-1}$ with power 7.29 in the 10-day spectrum, we now see a peak (A) at $9.43 \pm 0.05$ $y^{-1}$ with power 11.51 in the 5-day spectrum, where frequency error estimates indicate the frequency change that reduces the power by 2.3 in comparison with the value at the peak (corresponding approximately to a reduction in probability by a factor of 10 since $e^{-2.3} \approx 0.1$). This is virtually conclusive evidence that the primary peak is that at 9.43 $y^{-1}$, of which the peak at 26.57 $y^{-1}$ in the 10-day power spectrum (which has almost disappeared in the 5-day power spectrum) is an alias. If this interpretation is correct, we may expect to see an alias of the 9.43 $y^{-1}$ peak at the frequency 72.01 – 9.43 $y^{-1}$, i.e. at 62.58 $y^{-1}$, in the 5-day spectrum. We do indeed find a peak with power 5.36 at $62.56 \pm 0.08$ $y^{-1}$. Hence aliasing occurs in an analysis of the 5-day dataset also, but it is not as confusing as in the analysis of the 10-day data-set, owing to the bigger frequency separation.

There are two other notable peaks in the power spectrum: one (B) at $43.72 \pm 0.06$ $y^{-1}$ with power 9.83, and one (C) at $39.28 \pm 0.05$ $y^{-1}$ with power 8.91. Each of these three peaks is stronger in the 5-day spectrum than in the 10-day spectrum. We discuss the possible significance of these three peaks in the next section.

We now wish to incorporate information concerning the non-uniformity of the window functions, as represented by the list of mean live times. We therefore look for a simple model for the time window function $W_n$ that has mean value unity over the bin duration and for which

$$\frac{1}{t_{e,n} - t_{s,n}} \int_{t_{s,n}}^{t_{e,n}} dt\, W_n(t)\, t = t_{ml,n}, \tag{4}$$

where $t_{ml,n}$ is the "mean live time" for bin n. Clearly, there is no unique choice of $W_n(t)$ to meet these requirements. We here adopt the simple choice of adopting a constant value over the range $t_{s,n}$ to $t_{ml,n}$, and a different constant value over the range $t_{ml,n}$ to $t_{e,n}$. The appropriate values are found to be

$$\begin{aligned} W_n(t) = W_{l,n} &\equiv \frac{t_{e,n} - t_{ml,n}}{t_{ml,n} - t_{s,n}} \quad \text{for} \quad t_{s,n} < t < t_{ml,n}, \\ W_n(t) = W_{u,n} &\equiv \frac{t_{ml,n} - t_{s,n}}{t_{e,n} - t_{ml,n}} \quad \text{for} \quad t_{ml,n} < t < t_{e,n}, \end{aligned} \tag{5}$$

and we refer to this as a "double-boxcar" function. It seems likely that this is the choice for which the range of $W_n(t)$ is a minimum.

We find that the power spectrum computed by this double-boxcar method that incorporates the mean live times is indistinguishable from that computed from the single boxcar function that makes no reference to the mean live times. Clearly, the likelihood method is much less sensitive to non-uniformity of the data acquisition than is the Lomb-Scargle method. It therefore appears that if one uses the likelihood method there will be no need to take account of further details concerning the non-uniformity of the time window functions.

## 3. DISCUSSION

We now consider the possibility that the three peaks identified in the previous section are real, and seek possible interpretations. It is convenient to begin with peak C at frequency $\nu_C = 39.28 \pm 0.05$. If the oscillations in the neutrino flux are real, the modulation mechanism is likely to involve magnetic field. It is therefore interesting to compare the power spectrum of the neutrino measurements with the power spectrum of the magnetic field at Sun center for the period of operation of Super-Kamiokande. This is shown in Figure 3. We see that the fundamental and the "second" harmonic are quite weak, but we find significant power in the third through the seventh harmonics. When we combine these five power spectra by forming the "combined power statistic" (Sturrock 2003b, Sturrock & Wheatland 2003), we obtain the estimate $13.20 \pm 0.14$ y$^{-1}$ for the synodic rotation frequency (or $14.20 \pm 0.14$ y$^{-1}$ for the sidereal rotation frequency) of the magnetic field. This leads to the band $39.60 \pm 0.42$ y$^{-1}$ for the third harmonic of the synodic rotation frequency of the magnetic field. We see that peak C falls within this band. When we apply the shuffle test (Bahcall & Press 1991;

Sturrock, Walther, & Wheatland 1997), randomly re-assigning flux and error measurements (kept together) to time bins, we find that only 5 cases out of 1,000 yield a power 8.91 or larger in the search band $39.60 \pm 0.42$ y$^{-1}$.

If the solar neutrino measurements really are related to the solar magnetic field, we may expect to see some similarity between the time-evolution of the oscillations at the third harmonic of the rotation frequency. We have examined the relationship between the two power spectra by means of the minimum-power statistic (Sturrock 2003b; Sturrock & Wheatland 2003), and we find that the minimum of the two power spectra (as a function of frequency) has a maximum at the frequency 39.56. We have therefore formed from both datasets at this frequency the cumulative Rayleigh power defined by

$$S_R(v, t(n)) = \frac{1}{N} \sum_{k=1}^{n} \left| x_k e^{i 2\pi v t(k)} \right|^2 , \qquad (6)$$

where $x_k$ is the relevant variable normalized to have mean zero and standard deviation unity and N is the total number of data points. The result is shown in Figure 4. We note a clear similarity in the two curves, both powers growing from 1996 until 1998, then decreasing until about 2000.5 (near the onset of solar maximum), after which both increase sharply. This close correspondence in the cumulative Rayleigh power curves supports the conjecture that the neutrino flux and the solar magnetic field are related.

We now consider peak A at 9.43 y$^{-1}$ with power 11.51, and peak B at 43.72 y$^{-1}$ with power 9.83. We have found evidence (Sturrock 2002c) in Homestake (Davis & Cox 1991; Lande et al. 1992; Cleveland et al. 1995, 1998) and GALLEX-GNO (Anselmann et al. 1993, 1995; Hampel et al. 1996, 1997; Altmann et al. 2000) data for r-mode oscillations (Papaloizou & Pringle 1978; Provost, Berthomieu, & Rocca 1981; Saio 1982) that appear to be the origin of the well-known Rieger oscillation (Rieger et al. 1984) and of similar "Rieger-type" oscillations that have been discovered in recent years (Bai 1992, 1994, 2003; Kile & Cliver 1991). We therefore examine the possibility that peaks A and B may be related to such oscillations.

R-modes are retrograde waves that, in a uniform and rigidly rotating sphere, have frequencies

$$v(l, m, syn) = m(v_R - 1) - \frac{2 m v_R}{l(l+1)} \qquad (7)$$

as seen from Earth, where l and m are two of the usual spherical-harmonic indices, and $v_R$ is the sidereal rotation frequency. An observer co-rotating with the sphere would detect oscillations at the frequency

$$v(l, m, rot) = \frac{2 m v_R}{l(l+1)} . \qquad (8)$$

Since the mode frequency does not depend upon the radial index n, it seems likely that similar oscillations, with similar frequencies, could occur in thin spherical shells inside a radially stratified sphere.

R-mode oscillations may modulate the solar neutrino flux by moving magnetic regions in and out of the path of neutrinos propagating from the core to the Earth. The resulting oscillations in the neutrino flux would be formed from combinations of the frequency with which an r-mode oscillation intercepts the core-Earth line and the frequency with which a magnetic structure intercepts the core-Earth line. These combinations will have the form

$$\nu = \left| m(\nu_R - 1) - \frac{2m\nu_R}{l(l+1)} \pm m'(\nu_R - 1) \right| \tag{9}$$

where $m'$, the azimuthal index for the magnetic structure, may be different from that of the r-mode. For $m' = m$ and for the minus sign, equation (9) yields the frequency given by equation (8), and for the plus sign it yields

$$\nu(l, m, alias) = 2m(\nu_R - 1) - \frac{2m\nu_R}{l(l+1)} \ . \tag{10}$$

We may refer to this for convenience as an "alias" of the frequency given by equation (8). If we suspect that a peak in the power spectrum is due to a process leading to the frequency given by equation (8), it is reasonable to inspect the power spectrum to see if there is also evidence of the frequency given by equation (10).

For l = 2 and m = 2, and for the range of values of $\nu_R$ inferred from the magnetic-field data, we find that equation (8) leads us to expect an oscillation in the band $9.47 \pm 0.09$ $y^{-1}$, and equation (10) leads us to expect an oscillation in the band $43.33 \pm 0.47$ $y^{-1}$. We see that the peaks A and B fall within these two bands. On applying the shuffle test, we find only 6 cases out of 10,000 in which a peak with power 11.51 or larger occurs in the band $9.47 \pm 0.09$ $y^{-1}$, and only 5 cases out of 1,000 that yield a power 9.83 or larger in the band $43.33 \pm 0.47$ $y^{-1}$. We may estimate that we would find both a power 11.51 or larger in the band $9.47 \pm 0.09$ $y^{-1}$ and a power 9.83 or larger in the band $43.33 \pm 0.47$ $y^{-1}$ only about 3 times in $10^6$ trials.

If peaks A and B are indeed related, we may see some similarity between the time-evolution of the two oscillations. Figure 5 shows the cumulative Rayleigh powers of the two oscillations. We see they have a similar trend, and that some of the fine structure is quite similar, supporting the hypothesis that both oscillations are due to a single r-mode.

It appears that detailed analysis of the 5-day Super-Kamiokande data supports previous evidence from radiochemical data that the solar neutrino flux is variable. However, the case is persuasive only if one takes account of all available information. For instance, if we carry through a likelihood calculation that takes account of the timing and of the flux measurements, but ignores the experimental error estimates, the three peaks are still present, but with reduced power – peaks A, B and C then have powers 5.88, 7.20, and 7.19, respectively. If one then repeats the above shuffle test, one finds that each peak has a reduced significance of a few percent. In order to arrive at definitive conclusions concerning variability, it would be most helpful if data from both Cerenkov experiments (Super-Kamiokande and SNO) were to be packaged according to a standard protocol (such as one-day bins tied to Universal Time), and then made available to the interested community.

I wish to thank the Super-Kamiokande consortium for sharing their data, and Taeil Bai, David Caldwell, Alexander Kosovichev, Jeffrey Scargle, and Guenther Walther for their interest and helpful discussions. Thanks are due also to the referee, whose criticisms and suggestions were most helpful in revising the article. This work was supported by NSF grant AST-0097128.

FIGURES

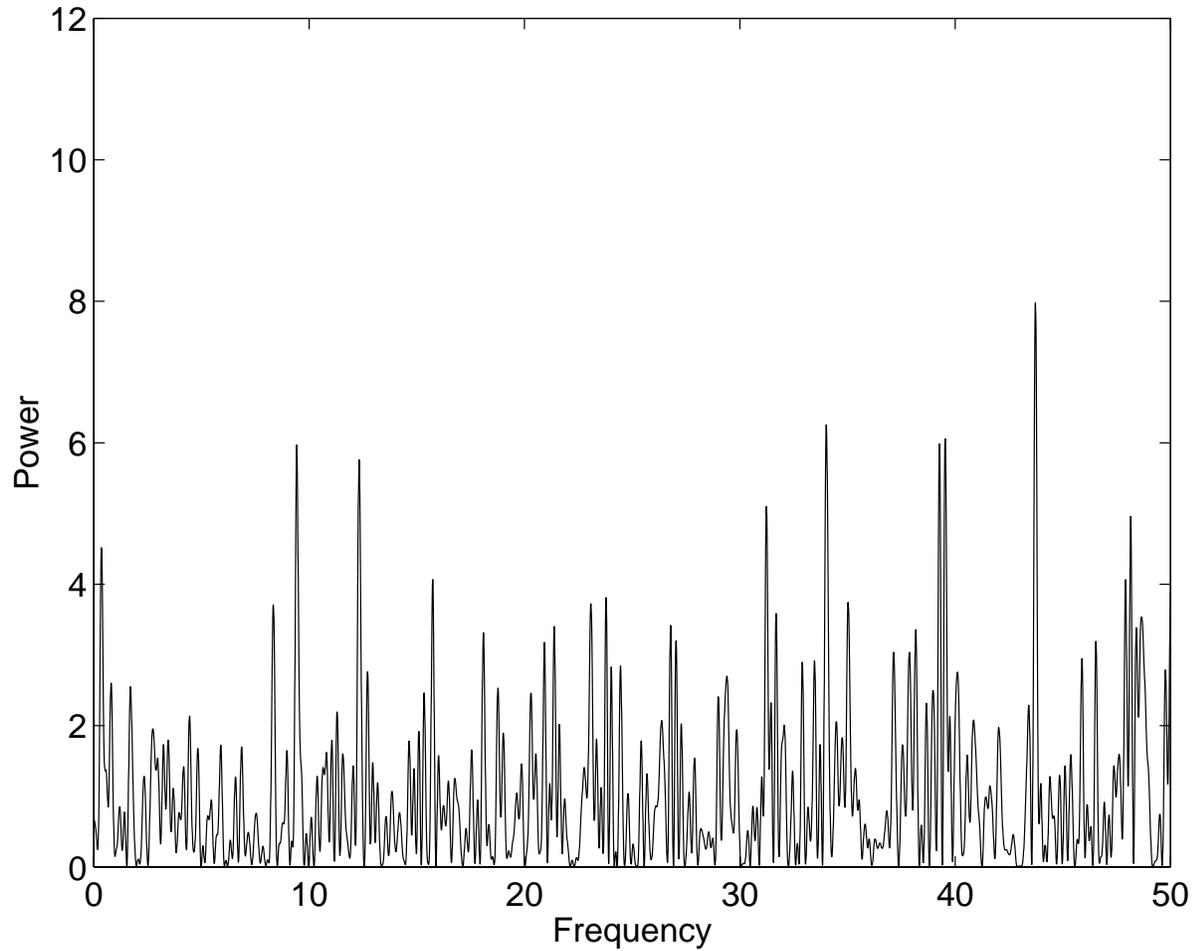

Figure 1. Power Spectrum of Super-Kamiokande 5-day data computed by the likelihood method, with a "delta-function" time window function and uniform weighting of measurements, to simulate the Lomb procedure.

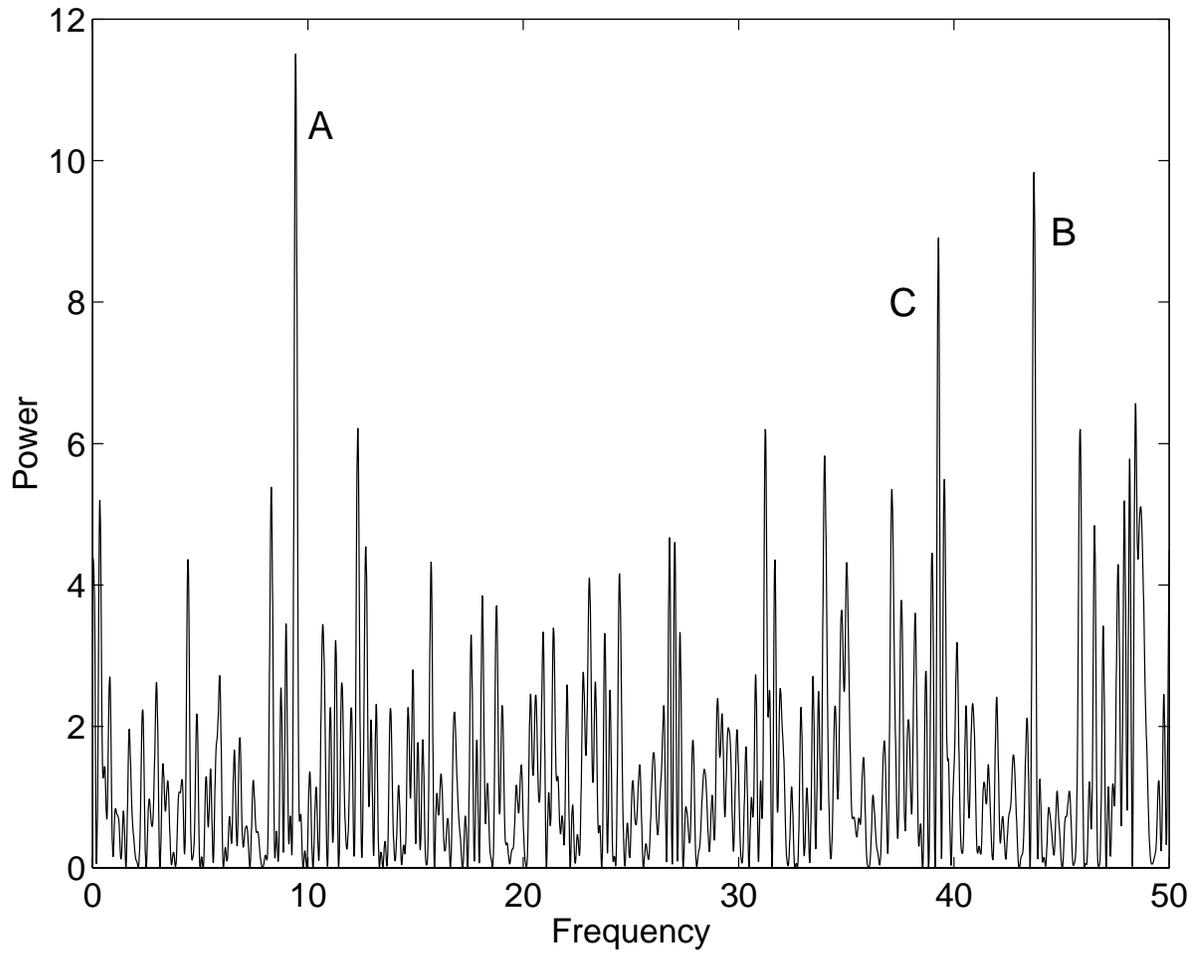

Figure 2. Power Spectrum of Super-Kamiokande 5-day data computed by the likelihood method, taking account of measurement error, start time, and end time, of each bin.

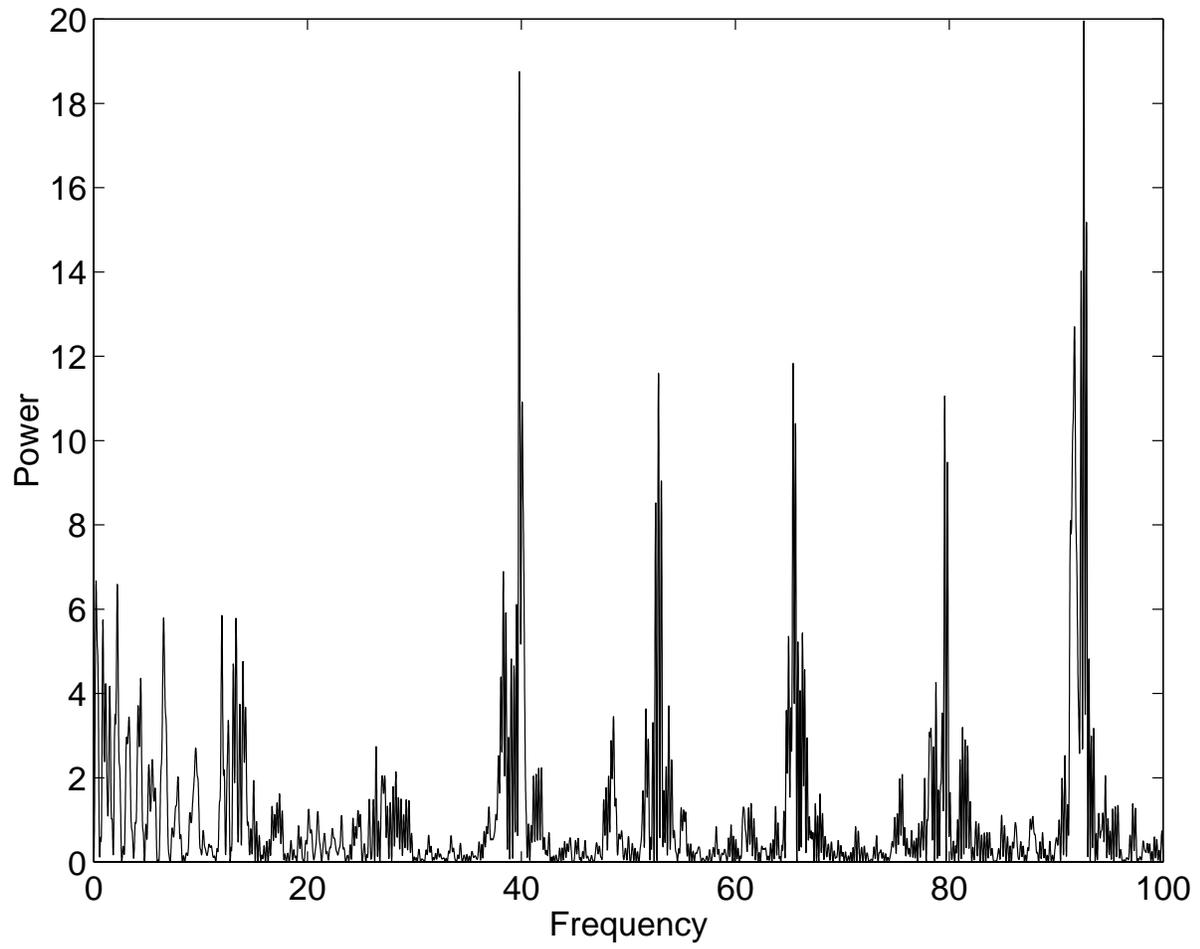

Figure 3. Power spectrum of the disk-center solar magnetic field for the interval of operation of Super-Kamiokande.

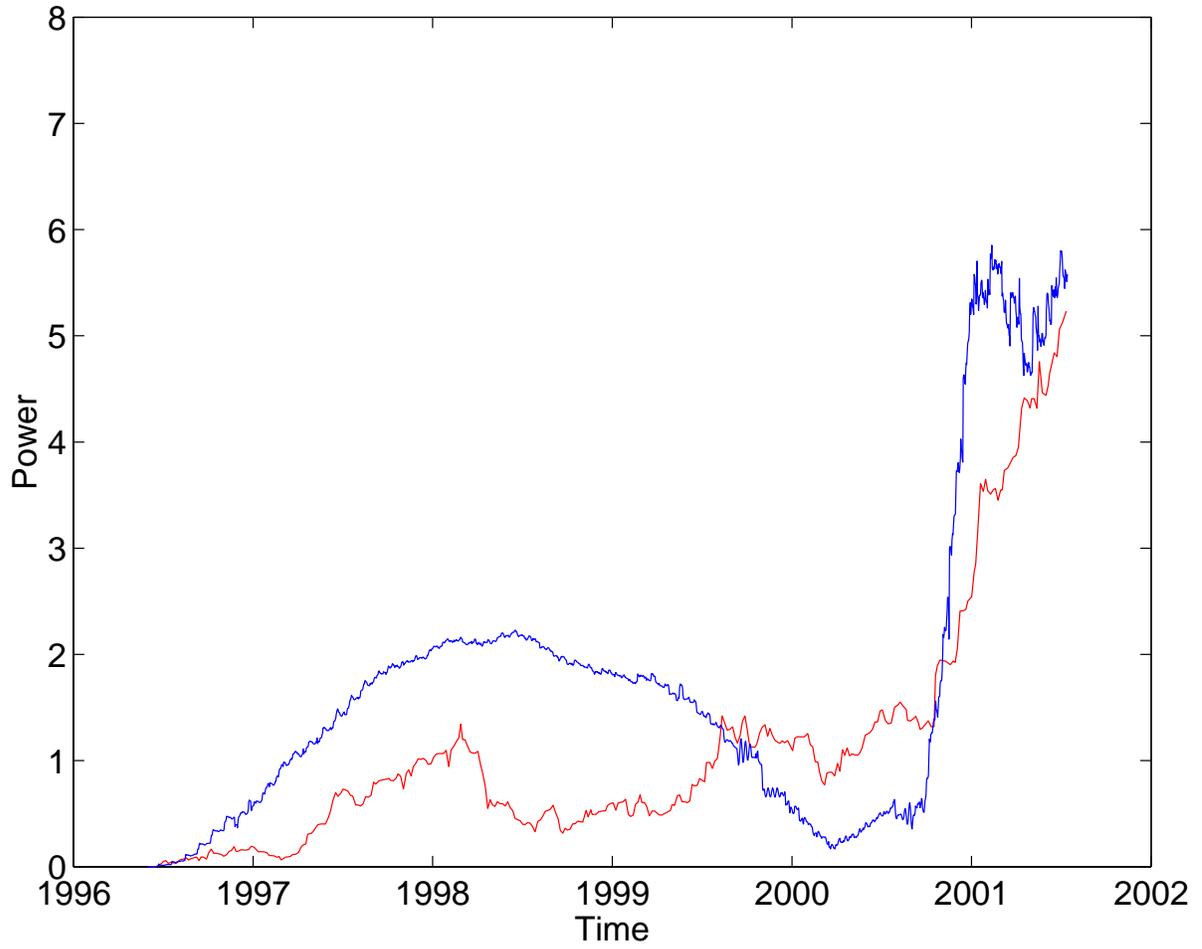

Figure 4. Cumulative Rayleigh power of disk-center magnetic field (blue) and 5-day Super-Kamiokande data (red) for frequency 39.56.

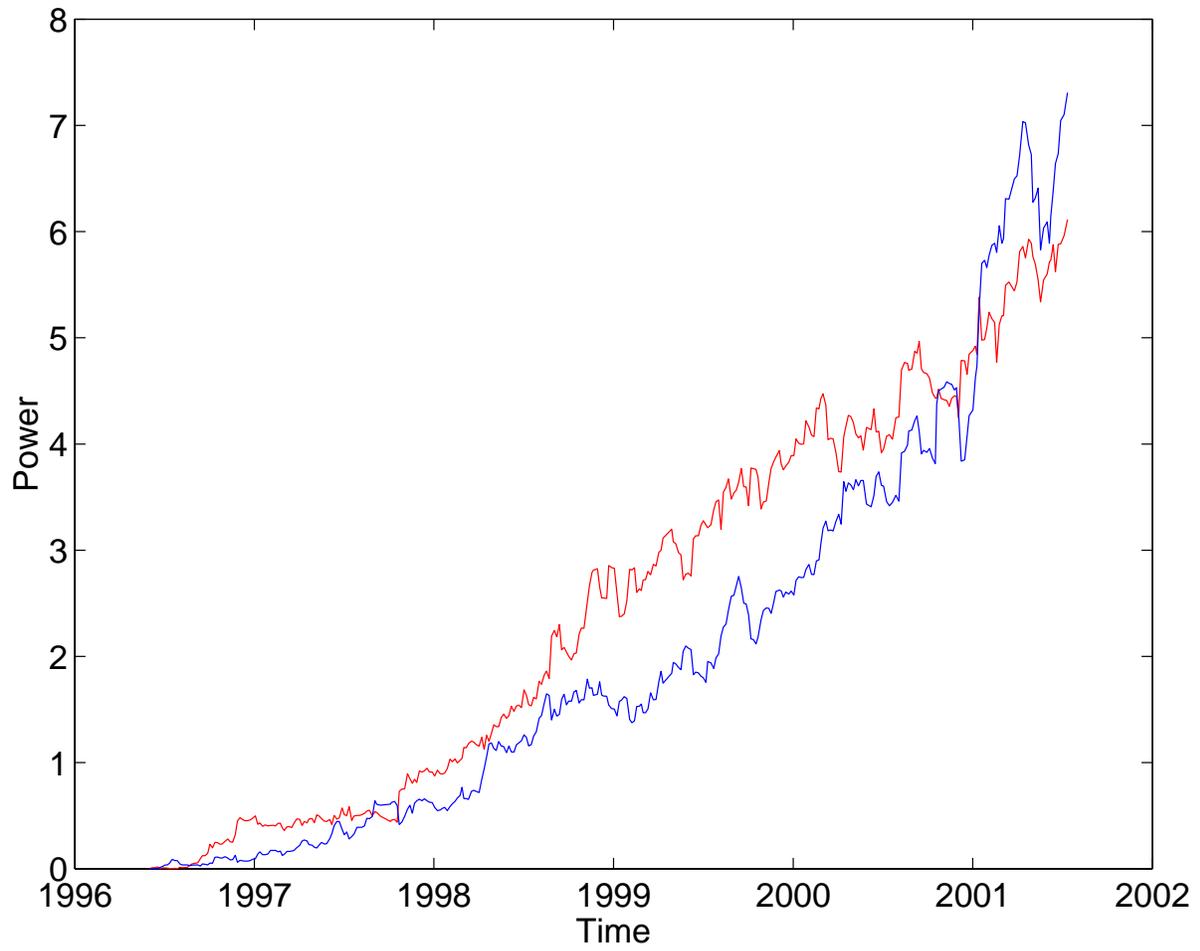

Figure 5. Cumulative Rayleigh powers of oscillations at 9.42 (red) and 43.72 (blue).